\documentclass[twocolumn,aps,prc,showpacs,preprintnumbers,amsmath,amssymb]{revtex4}
\usepackage{epsfig} 
\usepackage{dcolumn}

\begin{document}

\title{E0 emission in $\alpha$ + $^{12}$C fusion at astrophysical energies}
\author{G. Baur}
\affiliation{Forschungszentrum J\"ulich, Institut f\"ur Kernphysik, D-52425 J\"ulich, Germany}
\author{K. A. Snover}
\altaffiliation{electronic address: snover@u.washington.edu}
\affiliation{Center for Experimental Nuclear Physics and Astrophysics, University of  Washington,
Seattle,~Washington~98195}
\author{S. Typel}
\affiliation{GANIL, Bd.~Henri Becquerel, BP 55027, F-14076~Caen~Cedex~05, France}
\date{\today}

\begin{abstract}

We show that E0 emission in $\alpha$ + $^{12}$C fusion at astrophysically interesting energies is negligible compared to E1 and E2 emission.

\end{abstract}


\maketitle

The $^{12}$C + $\alpha \rightarrow ^{16}$O capture reaction, sometimes called the ``Holy Grail" of nuclear astrophysics, determines the ratio of  $^{16}$O to $^{12}$C at the end of helium burning in stars, which is very important for stellar evolution and nucleosynthesis~\cite{rolfs}.  Nucleosynthesis requires~\cite{weaver} a total S-factor for this reaction of about 170 keV b at a center-of-mass energy $E_{c.m.}$ = 0.3 MeV, the center of  the Gamow window.  The results of many experiments over more than 3 decades, extrapolated to the Gamow window, show that single-photon emission is dominated by E1 and E2 decay to the $^{16}$O ground-state, with approximately equal intensity and a combined S-factor S(0.3) approaching the value quoted above~\cite{hammer}.  The corresponding cross sections are $\sigma_{E1}$(0.3) $\approx \sigma_{E2}$(0.3) $\approx$ 1.4 x 10$^{-17}$ b.   

In this paper we examine the possible role of E0 emission, which has not, to our knowledge, been addressed previously.  We note that if E0 emission were important, it would have escaped observation in $^{12}$C + $\alpha \rightarrow ^{16}$O capture measurements since they were made by detecting the emitted  $\gamma$-rays, and the e$^+$e$^-$ pairs produced by E0 emission would not result in a sharp gamma line near the transition energy. 

First, we estimate the ratio of direct E0 and direct E2 emission,  following Snover and Hurd~\cite{snover}.  There, a general relation for direct E0 emission was derived, and for $^3$He + $^4$He fusion at low energies a simple relation was obtained for the direct cross section ratio $\sigma_{E0} / \sigma_{E2}$, which was shown to be negligibly small.  This occurs primarily because E0 emission is suppressed by an additional power of $\alpha$, the fine structure constant, relative to E2 emission.   

However, in $^{12}$C + $\alpha \rightarrow ^{16}$O$_{g.s.}$ there are several factors that enhance the relative importance of E0 emission: 1) E0 emission occurs by s-wave capture, whereas E1 and E2 emission arise from  p-wave and d-wave capture, respectively; 2) E1 emission is isospin-inhibited; and 3) the higher transition energy results in larger E0/E1 and E0/E2 phase-space factor ratios.  

In low-energy $^3$He + $^4$He fusion,  E0 and E2 direct capture occur between the same initial and final states (p-waves), and as a result the direct capture radial matrix elements cancel in the cross section ratio.   In $^{12}$C + $\alpha \rightarrow ^{16}$O$_{g.s.}$, however, the radial matrix elements are different since the initial states are different.  In analogy with Eq. 11 of ~\cite{snover} we obtain   

\begin{equation}
\frac{\sigma_{E0}}{\sigma_{E2}} = \frac{4\pi}{5} \frac{f_{E0}}{f_{E2}} 
\frac{|R_{00}|^2}{|R_{02}|^2},
\label{ratio}
\end{equation}
where $R_{l_f l_i}$ is the magnitude of the radial integral of $r^2$ between the initial continuum state with orbital angular momentum $l_i$ and the final bound state with $l_f = 0$.  

The quantities $f_{EL}$ are given by~\cite{snover}

\begin{equation}
f_{E0}(E) =  \frac{e^4}{27(\hbar c)^6}b(S)(E-2mc^2)^3(E+2mc^2)^2,
\label{fE0}
\end{equation}
and
\begin{equation}
f_{E2}(E) = \frac{4\pi e^2}{75(\hbar c)^5}E^5
\label{fE2}
\end{equation}
where $E = E_{c.m.} + Q$ is the transition energy, $Q$ = 7.16 MeV, 
\begin{equation}
b(S) = \frac{3\pi}{8}\left(1-\frac{S}{4}-\frac{S^2}{8}+\frac{S^3}{16}-\frac{S^4}{64}+\frac{5S^5}{512}\right)
\end{equation}
and $S = (E-2mc^2)/(E+2mc^2)$.   We estimate $|R_{00}|^2/|R_{02}|^2 = P_0/P_2 = 18$ at $E_{c.m.}$ = 0.3 MeV, where $P_{l_i}$ is the penetrability due to the Coulomb and angular momentum barriers evaluated at the radius $R$ = 1.3(A$_1^{1/3}$+A$_2^{1/3}$) fm = 5 fm.  This yields  4.3 x 10$^{-3}$ for the direct (i.e., nonresonant) E0/E2 cross section ratio at 0.3 MeV.

This estimate for $|R_{00}|^2/|R_{02}|^2$ assumes the capture takes place at the nuclear radius and is not affected by the nuclear interaction between ${}^{12}$C and the $\alpha$ particle in the continuum.  However, at low collision energies the effective radius may be larger, due to the importance of extranuclear capture, which would reduce $|R_{00}|^2/|R_{02}|^2$.  In addition, the total E2 capture cross section in the Gamow window is dominated by the tail of the subthreshold 6.92 MeV 2$^+$ state, and this effect is also not included above.  

We have improved on the above estimate by carrying out potential model calculations of E0 and E2 emission in $^{12}$C + $\alpha \rightarrow ^{16}$O$_{g.s.}$.    Using a real Woods-Saxon potential with radius parameter $r_0$ = 1.25 fm and diffuseness $a $= 0.65 fm, we find $V$ =  63.87 MeV to bind the $N$ = 2, $L$ = 0 $0_1^+$ ground-state at the measured energy.  Here $N$ and $L$ are determined from the relation $2N + L = \Sigma (2n_j + l_j)$ where $n_j$ and $l_j$ are the shell model quantum numbers of the 4 nucleons (0p or 1s0d) that make up the alpha particle state with quantum numbers $N,L$ in the $\alpha$-nucleus potential.   Since the 6.05 MeV $0_2^+$ state and the 6.92 MeV $2_1^+$ state are members of the same 4p-4h rotational band, with the particles in the 1s0d shell, they should both have $2N + L$ =8 and hence $N$ = 4 for the $0_2^+$ state and 3 for the $2_1^+$ state.  We find $V$ = 122.74 MeV (122.03 MeV) to bind the $0_2^+$ ($2_1^+$) states with these node numbers at the correct energy, and thus we use $V(l_i=0)$ = 122.74 MeV and  $V(l_i=2)$ = 122.03 MeV for the $l_i$ = 0 and 2 scattering states, respectively, and $V(l_f=0)$ = 63.87 MeV for the final state.  We note that these scattering potentials are similar to the real Woods-Saxon potential that fits the rainbow scattering region in intermediate energy $\alpha$ - $^{12}$C elastic scattering~\cite{goldberg}.

\begin{figure}
\includegraphics[width=0.48\textwidth]{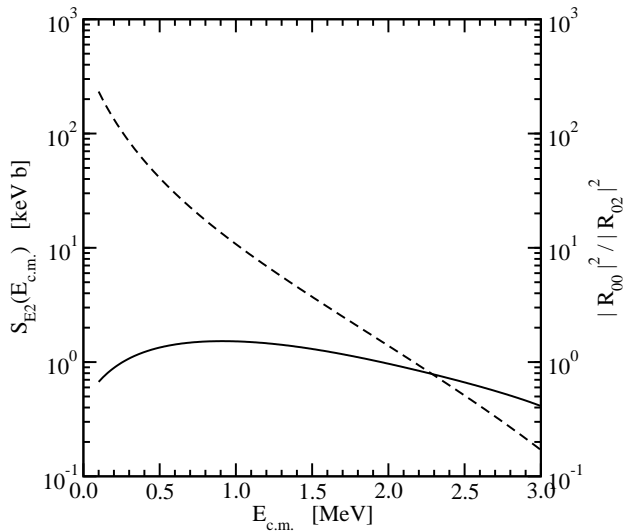}
\caption{Dashed curve and left scale: E2 S-factor;  solid curve and right scale:  E0/E2 radial matrix element ratio; vs. $E_{c.m.}$.}
\label{figure}
\end{figure}

With these potentials, we obtain the E2 S-factor shown in Fig.~\ref{figure}.  This curve is within a factor of 2 of the measured E2 S-factors below E$_{c.m.}$ = 2 MeV, and has $S_{E2}$(0.3) = 85 keV b, in agreement with the value 81 $\pm$ 22 keV b obtained by Hammer et al.~\cite{hammer} from an extrapolated R-matrix fit to E2 data (other modern E2 fits that we are aware of yield $S_{E2}$(0.3) values within a factor of 2 of these values).  

Our potential model results for $|R_{00}|^2/|R_{02}|^2$ are also shown in Fig.~\ref{figure}.  We obtain a value of 1.1 for the ratio at 0.3 MeV.  This may be compared to the value 3.2 calculated with a pure $l_i$ = 0 Coulomb scattering wave, indicating that the interior and exterior contributions to the E0 matrix element interfere destructively.  A calculation with $V(l_i=0)$ = 122.03 MeV, which artificially enhances the contribution of the subthreshold $0_2^+$ state by moving it 0.2 MeV closer to threshold,  yields a ratio of  2.0 at 0.3 MeV.  With $|R_{00}|^2/|R_{02}|^2$ = 1.1, our calculated E0/E2 cross section ratio is 2.6 x 10$^{-4}$.   Taking S$_{E2}$(0.3) = 80 keV b, this corresponds to 

\begin{equation}
S_{E0}(0.3) = 0.02 \mbox{ keV b}.
\label{se0}
\end{equation}

\begin{table}
\caption{0$^+$ resonance tail and potential model contributions to E0 emission at 0.3 MeV.}
\label{data}
\begin{ruledtabular}
\begin{tabular}{cclrr}
E$_x$(MeV) & $\theta_{\alpha_0}^2$ & M(fm$^2$)\footnotemark[1] & $\sigma_{E0}$(0.3)(b)  & Ratio\footnotemark[2] \\
\hline 6.05 & $\leq$ 0.7\footnotemark[3] &  3.55 & $\leq$ 1.6x10$^{-21}$ & $\leq$ 1.2x10$^{-4}$  \\
12.05 & 0.0036\footnotemark[1]$^,$\footnotemark[4] & 4.03 & 1.0x10$^{-25}$ &7.8x10$^{-9}$  \\
14.03 & 0.031\footnotemark[1]$^,$\footnotemark[4] & 3.3 & 2.9x10$^{-24}$ &2.2x10$^{-7}$  \\
25 & $\leq$ 1.0  & 9.0\footnotemark[5] & $\leq$ 1.0x10$^{-22}$ & $\leq$ 7.3x10$^{-6}$  \\
\hline
potential model&  &     &  &  2.6x10$^{-4}$ \\ 
\end{tabular}
\end{ruledtabular}
\footnotetext[1]{monopole decay matrix element~\protect\cite{tilley}.}
\footnotetext[2]{ $\sigma_{E0}$/$\sigma_{E2}$(total) at 0.3 MeV, where $\sigma_{E2}$(total) = 1.4x10$^{-17}$ b.}
\footnotetext[3]{see e.g. Table IV of~\protect\cite{desc}.}
\footnotetext[4]{ $\Gamma_{\alpha_0}/(2P_0 \gamma_{W.L.}^2)$ where $\gamma_{W.L.}^2 $=  $3\hbar^2/(2\mu a^2)$ = 0.82 MeV.}
\footnotetext[5]{$M^2 = (0.83)8\hbar^2<r^2>_{prot}/(E_xM_n)$ where $<r^2>_{prot}$ = 7.34 fm$^2$~\protect\cite{tilley} and $M_n$ = nucleon mass.}
\end{table}

Tails of higher lying 0$^+$ resonances may also contribute to the E0 cross section.
In Table \ \ref{data} we show the 0$^+$ excited states of $^{16}$O with known ground-state monopole decay strengths~\cite{tilley}.  Also shown for each state is the  reduced $\alpha_0$ width in units of the Wigner Limit, the monopole decay matrix element, the E0 cross section  at 0.3 MeV based on a Breit-Wigner extrapolation using the s-wave penetrability, and the ratio of the E0 cross section to the total E2 cross section at 0.3 MeV.  We show an estimate for the 6.05 MeV $0_2^+$ state for completeness, even though its effect on the cross section is included in the potential model calculations.  We also show an upper limit for the contribution of the tail of an isoscalar Giant Monopole Resonance located at E$_x$ = 25 MeV with 83\% of the isoscalar energy weighted sum rule~\cite{bohr} (the remaining 17\% resides in the other 0$^+$ states shown in Table \ \ref{data}).   None of the resonance tail contributions from states above 6.05 MeV are significant compared to the E0 cross section calculated in the potential model.   

E0 emission to excited final states in $^{16}$O is negligible due to the small phase space factor.  Hence our best estimate for the E0 contribution to the astrophysical S-factor for  $^{12}$C + $\alpha$ capture is given by Eq.~\ref{se0} above.

Two-photon emission is also negligible, based on the measured branching ratio for this process in the decay of  the 6.05 MeV 0$^+$ state~\cite{watson}.  We conclude that electromagnetic processes other than single-photon emission do not contribute significantly to the astrophysical rate for  $^{12}$C + $\alpha$ fusion.

We thank C. Rolfs for bringing this problem to our attention, and the U.S. DOE, Grant \# DE-FG02-97ER41020 for financial support.


\begin{thebibliography}{}
\bibitem{rolfs} See e.g. C. E. Rolfs and W. E. Rodney, Cauldrons in the Cosmos, (U. Chicago Press, 1988).
\bibitem{weaver} T. A. Weaver and S. E. Woosley, Phys. Rep. {\bf 227},  65 (1993).  See also T. Rauscher et al., Astrophys. J. {\bf 576}, 323 (2002); S. E. Woosley et al., Nucl. Phys. A {\bf 718}, 3c (2003).
\bibitem{hammer} J. W. Hammer et al., Nucl. Phys A {\bf 768}, 353c (2005).
\bibitem{snover} K. A. Snover and A. E. Hurd, Phys. Rev. C {\bf 67}, 055801 (2003).
\bibitem{desc} P. Descouvemont, D. Baye and P.-H. Heenen, Nucl. Phys. A {\bf 430}, 426 (1984).
\bibitem{goldberg} D. A. Goldberg, Phys. Lett. {\bf 55} B, 59 (1975).
\bibitem{tilley} D. R. Tilley, H. R. Weller and C. M. Cheves, Nucl Phys. A {\bf 564}, 1 (1993).
\bibitem{bohr} A. Bohr and B. R. Mottelson,  Nuclear Structure (Benjamin, Reading MA, 1975) Vol. II, Eqs. (6-178).  The IS sum rule is 1/2 of the combined IS + IV sum rule given here.
\bibitem{watson} B. A. Watson et al., Phys. Rev. Lett. {\bf 35}, 1333 (1975).
\end{thebibliography}
\end{document}